\documentstyle[twocolumn,aas2pp4]{article}
\def\stacksymbols #1#2#3#4{\def\theguybelow{#2}
        \def\verticalposition{\lower#3pt}
        \def\spacingwithinsymbol{\baselineskip0pt\lineskip#4pt}
        \mathrel{\mathpalette\intermediary#1}}
\def\intermediary #1#2{\verticalposition\vbox{\spacingwithinsymbol
        \everycr={}\tabskip0pt
        \halign{$\mathsurround0pt#1\hfil##\hfil$\crcr#2\crcr
                \theguybelow\crcr}}}
\def\lta{\stacksymbols{<}{\sim}{2.5}{.2}}
\def\gta{\stacksymbols{>}{\sim}{3}{.5}}

\begin{document}

\title{Hot Gas and Halos in Elliptical Galaxies}
\author{William G. Mathews}
\affil{University of California Observatories/Lick Observatory,
Board of Studies in Astronomy and Astrophysics,
University of California, Santa Cruz, CA 95064}

\author{Fabrizio Brighenti}
\affil{University of California Observatories/Lick Observatory,
Board of Studies in Astronomy and Astrophysics,
University of California, Santa Cruz, CA 95064}
\affil{Dipartimento di Astronomia,
Universit\`a di Bologna,
via Zamboni 33, Bologna 40126, Italy}

\begin{abstract}
We review recent progress in understanding the evolution
of hot interstellar gas in the halos of elliptical galaxies.
The enormous spread in x-ray luminosity
$L_x$ for galaxies of similar $L_B$
is driven by non-homologous variations in
the physical size of the hot interstellar medium,
$L_x \propto L_B^2 (r_{ex}/r_e)^{0.6}$, where $r_{ex}$ and
$r_e$ are the half-luminosity radii in x-rays and optical radiation.
This relation may have been established as the ellipticals
formed in small groups of galaxies.
By combining ROSAT with older {\it Einstein} data we
derive the distribution of
total mass $M_{tot}(r)$ in NGC 4472 which,
for $0.1 \lta r/r_e \lta 1$, is identical to the
expected stellar mass $M_*(r)$.
This means that
stellar mass to light ratios can be determined
from x-ray observations!
It also means that the widely used ``mass dropout'' assumption
must be incorrect in this important part of the cooling flow.
Recent ROSAT determinations of temperature profiles
in ellipticals show a curious maximum in the gas temperature at
$\sim 3~r_e$ and global mean gas temperatures that are
about twice the virial temperature of the stars.
These results are totally unlike those predicted by standard
models of galactic cooling flows.
However, cooling flow solutions can be brought into 
agreement with observed interstellar
temperature and density profiles if they begin 
with an additional
massive component of ``circumgalactic'' gas, assumed 
to fill the
outer galactic halos beyond most of the stars.
This old hot gas, first heated during the epoch of galaxy
formation, continues to flow into the stellar parts of
ellipticals today, combining with gas expelled from
evolving stars.
The dual origin of hot interstellar gas further complicates
recent discussions of abundances in the
hot interstellar gas.
\end{abstract}


\section{Introduction}

Gravitational confinement of hot, x-ray emitting 
interstellar gas provided the earliest evidence for 
massive halos in early type galaxies
(Bachall \& Sarazin 1977; Mathews 1978;
Forman, Jones, \& Tucker 1985).
The ratio of interstellar to stellar baryonic components,
3 - 10 percent,
is similar to that in typical spiral galaxies.
Dynamically this gas must be quiescent, near hydrostatic equilibrium.
If the observed interstellar 
gas were free-falling in the galactic potential it could not 
be replenished fast enough from stellar mass loss to maintain
the gas mass observed.
If the hot gas were participating in a supersonic 
galactic wind, the gas density and x-ray luminosity 
would be very much less than observed.
Therefore, the hot gas must be close to static equilibrium. 
Hydrostatic equilibrium 
is also supported by the short sound-crossing time 
to the effective radius, $t_{sc} \sim 10^8$ yrs.
The total mass within 
radius $r$ can be determined from the condition for 
hydrostatic equilibrium,
\begin{displaymath}{
M_{tot}(r) = - {k T(r) r \over G \mu m_p }
\left( {d \log \rho \over d \log r} + {d \log T \over d \log r}
+ {P_{mag} \over P} { d \log P_{mag} \over d \log r}
\right).
}
\end{displaymath}
The total mass is sensitive to the leading coefficient 
which involves the gas temperature $T(r)$ determined
from the x-ray spectrum.
The logarithmic derivatives describe the 
radial variation of the total pressure. 
The (negative) gas density derivative is generally 
the largest and can be rather accurately determined from the 
x-ray surface brightness distribution.
Typical temperature gradients are smaller and 
only weakly influence $M_{tot}(r)$.
Very little is known 
of the variation of magnetic pressure $P_{mag} = B^2/8\pi$.
However, since $(P_{mag}/P) ( d \log P_{mag} / d \log
r)$ is expected to be negative, if this term were important and
not considered in the hydrostatic equilibrium, then 
$M_{tot}$ would be {\it underestimated}. 
Determinations of the total mass $M_{tot}(r)$ in elliptical
galaxies clearly indicate  
the universal presence of dark matter halos.
For those ellipticals that are members of rich clusters
having hot, high pressure cluster gas,
the partial confinement of interstellar gas by the ambient 
cluster gas cannot simulate the effect of dark halos;
$M_{tot}(r)$ can still be determined from hydrostatic equilibrium
in this case too provided the 
logarithmic gradients in the equation above can be 
accurately determined.

While much was learned about elliptical galaxies 
from {\it Einstein} observations, more recent high quality 
x-ray observations with ROSAT and ASCA 
have revealed unanticipated aspects of the hot interstellar gas 
that are inconsistent with older theoretical models.
In the following brief review we discuss some of these
recent developments and their possible implications. 
We show that the sizes of the 
x-ray images, now well-determined for about a dozen
ellipticals, account for the strong scatter in 
x-ray luminosity among ellipticals having similar optical 
luminosity.
We have also realized that interior to an effective radius 
in NGC 4472 and 4649 the hot gas pressure 
is balanced by the {\it stellar} potential, with little 
or no influence from dark matter; this allows a determination 
of the stellar mass to 
light ratio directly from x-ray data.
Of particular interest are the strange maxima in the 
interstellar gas temperature, peaking near $r \sim 3~r_e$, 
and the generally high value of the mean gas temperature in 
ellipticals, about 
twice the equivalent temperature of orbiting 
stars $T_*$.
We show below that this temperature profile cannot be produced
by normal mass loss from evolving galactic stars, instead
most of the hot gas currently observed in elliptical galaxies
may have been stored in the outer dark halo regions since 
the formation of these galaxies by mergers and tidal truncations
in small groups of galaxies.
But many details are still incompletely understood.
It remains a mystery, for example, why
few if any x-ray images show the isophotal flattening
expected from the global rotation of 
galactic stars as observed in 
essentially all ellipticals.

We begin with a brief review of the nature of 
galactic cooling flows.

\section{Cooling Flows in Elliptical Galaxies}

In the standard explanation, to be modified below,
the hot interstellar gas in ellipticals is the 
accumulation of normal mass loss from a dominant 
population of old galactic stars.
Many lines of evidence suggest that the stars in large 
ellipticals are very old, formed in nearly coeval 
bursts of star formation (Bender 1997).
The gaseous envelopes expelled from these normally 
evolving stars  
collide with the ambient gas, shock, and dissipate the
orbital energy of the parent stars.
Evidently, the efficiency of this dissipation accounts for 
the high gas temperature in ellipticals 
and explains why it is roughly comparable to the virial
temperature of stellar orbits $T_* \sim 10^7$ K. 
Although the hot gas could be 
heated further by Type Ia supernovae, 
this cannot be very important since the iron abundance
in the interstellar gas would greatly exceed that observed.
If the hot gas is in quiescent equilibrium, it is easy to 
show under these assumptions that 
the x-ray luminosity should vary roughly as the square of
the optical luminosity, $L_x \propto L_B^2$ (e.g Tsai \& Mathews 1995)
and this 
dependence has been observed (e.g. Eskridge, Fabbiano \& Kim 1995)
but with very large cosmic scatter.

Why is hot gas in ellipticals referred to as ``cooling flows''?
The observed x-rays are clear evidence that the hot gas is losing 
energy.
However, as energy is lost by radiation, the gas sinks deeper
into the galactic potential and receives additional energy 
by compression in the atmosphere of hot gas.
The net result, quite ironically, is that the gas 
in the ``cooling flow'' doesn't 
cool until its density is so high that $Pdv$ heating is 
no longer able to compensate radiative losses.
Under these circumstances the gas temperature profile 
in a galactic ``cooling flow'' is 
nearly constant with galactic radius.
Since all ellipticals exhibit some amount of rotation,
the final stage of cooling is likely to be into a disk
rotating at the local circular velocity.
In the absence of strong supernova heating or other disturbances, 
the cooling flow velocity is negative and highly subsonic 
throughout the cooling flow except immediately before the
final cooling catastrophe where the flow passes through a 
sonic surface and shocks against the cooled gas already present.
Since the stellar mass loss varies with time (see below), the flow 
continuously evolves and never reaches steady state.

\section{X-ray and Optical Image Sizes Correlate with $L_x/L_B$}

In spite of the overall correlation $L_x \propto L_B^2$,
ellipticals with similar $L_B$ can have x-ray luminosities $L_x$
that range over factors of 30 - 100.
This large scatter has been discussed for many years but 
no intrinsic property of the galaxies had been found
that correlated with residuals from the $L_x,L_B$ correlation.
However, we have recently shown that these residuals are related 
to the sizes of the x-ray images.
As the number of ellipticals that are spatially resolved 
in x-radiation slowly increases, the wide disparity in 
their x-ray image size $r_{ex}$ has become evident 
(Loewenstein 1996).
Ellipticals that are centrally dominant in small galaxy groups 
-- NGC 5044 or 1399 -- have enormous clouds of hot gas 
extending far beyond their optical images.
Other less endowed ellipticals -- NGC 4374 or 4649 -- have
x-ray images that are truncated at or near their optical
radius $r_e$.
Ellipticals are thought to have formed by galactic 
mergers in small galaxy groups.
Their close dynamical proximity in group environments 
also promotes tidal exchanges of halo material (Merritt 1985;
Bode et al. 1994) and probably also of hot gas.
Motivated by this conjecture,
Mathews \& Brighenti (1997b) plotted $r_{ex}/r_e$ against 
residuals in the $L_x,L_B$ plot and found a strong correlation.
A distance-independent representation of this correlation,
$(L_x/L_B) \propto (r_{ex}/r_e)^{0.6 \pm 0.3}$,
is shown in Figure 1.
Although the significance of this result is not fully understood,
it is likely that the vast range of x-ray sizes
is related to disparities in the final allocation of halo 
material resulting from mass-exchange interactions
in group environments.
If this explanation is correct, 
information about hot gas production in the 
early history of ellipticals 
must still be retained in currently observed ellipticals.
We think that this is indeed the case.

\section{Evolutionary Models of Galactic Cooling Flows}

To allow for the stellar origin of the gas, 
the standard equations of gas dynamics 
expressing  conservation of mass, momentum and energy 
must be embellished with source and sink terms representing
the input of gas from orbiting stars, the possible heating
by Type Ia supernovae, and cooling of the gas by radiation 
losses (e.g. Brighenti \& Mathews 1996).
The steadily decreasing 
rate that mass is supplied from an old single-burst
stellar population varies as $\alpha_*(t) \rho_*(r)$
where $\alpha_*(t) = \alpha(t_n) (t/t_n)^{-1.3}$ s$^{-1}$
(Mathews 1989) and $t_n = 13 - 15$ Gyrs is the current age
of the population.
The value of $\alpha_*(t_n)$ corresponds to about 1.5 
$M_{\odot}$ yr$^{-1}$ for a galaxy having total stellar mass 
$M_{*t} = 10^{12}$ $M_{\odot}$.
The gas moves in a galactic potential described by 
a King or de Vaucouleurs stellar system immersed in 
a massive dark halo.
Since intercluster gas in rich clusters is rich 
in iron ($[z/z_{\odot}] \approx 0.3$), it has usually 
been assumed that galactic winds driven by Type II 
supernovae were common early in the 
history of elliptical galaxies (David et al. 1990; 1991).
For this reason evolutionary cooling flow calculations
often begin with an essentially gas-free galaxy
at some early time $t \approx 1$ Gyr when it is assumed
that that galactic winds subsided; thereafter the interstellar 
gas is provided by stellar mass loss.

The results of a typical evolutionary calculation of this
sort for a non-rotating E0 galaxy on the fundamental plane 
are shown in Figure 2 (from Brighenti \& Mathews 1996).
The gas temperature variation $T(r)$ follows 
the stellar temperature
$T_*(r)$ which is found by solving the Jeans equation 
in the combined stellar-dark halo potential.
Of particular interest is the small but nevertheless significant
difference shown in Figure 2 
between the projected soft x-ray surface brightness distribution 
$\Sigma_x(R)$ (in the {\it Einstein} band) 
and the observed x-ray profile $\Sigma_{x,obs}(R)$ which 
for some well-observed galaxies is very similar to the 
stellar surface brightness distribution 
$\Sigma_*(R)$ (Trinchieri et al. 1986).
The computed x-ray surface brightness is too centrally
peaked.
The traditional means of correcting this discrepancy
has been to remove hot gas from the flow, removing more 
gas closer to the galactic center (Fabian \& Nulsen 1977; Stewart et
al. 1984; White \& Sarazin 1987a, 1987b, 1988; 
Thomas et al. 1987; Sarazin \& Ashe 1989; Bertin \& Toniazzo 1995).
This ``mass dropout'' 
introduces an arbitrary adjustable function of 
galactic radius into the theoretical model, allowing 
$\Sigma_x(R) \propto \Sigma_{x,obs}(R)$ at least for 
small $R$.
The physical justification for this approach has been
the notion that thermal instabilities rapidly 
remove hot gas throughout cooling flows,
not just at the central regions.
Only low mass stars are assumed to form in these 
unstable condensations
since a normal population of 
young stars is inconsistent with the red optical spectrum
observed in most ellipticals.
Although the resulting cooling flow models are improved with 
the ``mass dropout'' assumption, dropout has not been supported
by detailed hydrodynamical calculations. 
Malagoli et al. (1990) and Hattori \& Habe (1990) showed that 
thermally unstable regions undergo motions due to
buoyant forces that introduce shear and Rayleigh-Taylor 
instabilities, effectively destroying the instability.
Moreover, high quality x-ray observations in a few bright
ellipticals are inconsistent with the ``mass dropout''
assumption (see below).
The total x-ray luminosity $L_x(t_n)$ for the model shown in
Figure 2 falls among the observed data in the
$L_x,L_B$ plot but this can hardly be regarded as a 
verification of the theory since the scatter in this
diagram is so large.

More realistic models than that shown in Figure 2 
must consider the influence
of galactic rotation on the evolution of hot interstellar 
gas.
Rotation is also expected to lessen the slope of 
the computed x-ray surface brightness distribution 
since the gas cools toward a disk configuration
and does not flow entirely to the center of the galactic
potential.
Even for the most luminous ellipticals, where galactic 
flattening of the stellar component 
is due to anisotropic stellar orbits,
some significant rotation is always observed,
$v_{*,rot} \sin i \approx 50 - 100$ kpc.
Less luminous ellipticals rotate even faster, 
having rotationally flattened optical images, and 
typically have stellar disks 
and evidence for youthful stars (Davies 1997; 
Scorza and Bender 1995).
Brighenti \& Mathews (1996; 1997b) have computed 
the evolution of rotating cooling flows in both slowly 
and rapidly rotating ellipticals.
$L_x$ is found to decrease markedly with increasing rotation.
A large disk $r_{disk} \sim r_e$ 
of hot gas forms as shown in Figure 3, 
even in slowly rotating ellipticals.
If such cooling gaseous disks 
formed into stars less massive than 
8 $M_{\odot}$ (since SNII are not observed),
stellar disks like those observed 
in disky, rapidly rotating ellipticals can be created 
from cooling flow gas.
But this theoretical model may be incomplete since
there is no evidence for either x-ray or stellar 
disks in massive ellipticals.

\section{Recent ROSAT and ASCA Observations}

Recent observations of elliptical galaxies 
with the ROSAT and ASCA satellites have
provided qualitatively new information about the iron abundance, 
gas temperature and density distributions in galactic 
cooling flows.
As a consequence, a much different
interpretation has emerged 
in which most of the hot gas is not supplied 
by stellar mass loss but instead is 
left over from the epoch of galaxy 
formation.

In principle the iron abundance in the hot gas can 
be found from observations of the complex of FeL lines
at $\sim 1$ keV.
In practice, however, there is a wide range of 
observed values 
depending on the data reduction procedure 
and instrument used.
For example, the iron abundance for NGC 4472 quoted 
by Awaki et al. (1994), $z_{Fe}/z_{Fe \odot} = 0.63$, is about 
half that of Buote \& Fabian (1997), $z_{Fe}/z_{Fe \odot} = 1.18$, 
although both groups were using ASCA data.
Some, but not all, of this variation is due to
significant differences in the adopted 
iron abundance for the sun
or solar system (Ishimaru \& Arimoto 1997).
Much discussion has focused on the low values
of the gas iron abundance relative to the iron abundance 
observed in the stellar spectrum
(Arimoto et al 1997); questions have been raised
about the accuracy of the FeL iron determination (Renzini 1997).
The abundance 
discrepancy is so large that the iron contribution 
from SNIa to the interstellar gas 
must be very low indeed (e.g. Loewenstein et al. 1994).
There is also some considerable uncertainty in the stellar 
iron abundances, particularly since it has been realized
that the relative abundances are non-solar.
The most recent contribution to this discussion, 
an extensive re-observation of NGC 4636 with ASCA, 
has resulted in iron abundances that are consistent
with stellar values (Matsushita et al. 1997).
Nevertheless, 
in view of the many uncertainties and the prospect that 
much of the hot gas does not come from galactic stars, 
we shall not discuss comparisons of computed and observed
cooling flow abundances here.

One of the least expected recent findings has been the high 
gas temperatures and strange 
radial temperature profiles in galactic cooling flows.
If the gas temperature $T$ is plotted as function of
normalized radius $r/r_e$ for the six bright, well-resolved 
ellipticals observed by ROSAT --
NGC 507, 1399, 4472, 4636, 4649, and 5044 --
a consistent non-isothermal temperature profile is apparent.
In each case the gas temperature is close to $T_*$ near the
center then rises to a maximum of 
about 1.5 - 2 $T_*$ at $\sim 3~r_e$ and slowly decreases
at $r > 3~r_e$ (Brighenti \& Mathews 1997a).
This $T(r)$ has been confirmed by ASCA observations 
for several of these ellipticals.
The region of positive 
$dT/dr$ within $3~r_e \sim 15 - 30$ kpc
is not a natural result of conventional theoretical models
(cf. Figure 1).
In these models 
the dominant stellar potential in this
region of the galaxy provides enough compressive heating
to keep the inflowing gas nearly isothermal.
In addition, 
Davis \& White (1996) have shown for a larger sample of
ellipticals that the average gas temperature is
$\langle T \rangle \approx 1.5 - 2 T_*$, consistent
with the peaking temperature profile of the six resolved
galaxies.
In the absence of SNIa heating, 
gas expelled from stars cannot have mean temperatures
greater than $T_*$ which characterizes the potential well
within the stellar system.
However, gas with temperature $T \sim 2 T_*$ can be 
bound to the outer dark halos.

\section{A Model for NGC 4472 Including Circumgalactic Gas}

Inspired by the serious qualitative differences 
between the observed gas 
temperature profiles and those expected from simple cooling 
flow models like that shown in Figure 1,
we decided to explore new types of evolutionary models using 
the observed properties of NGC 4472 as a guide.
While NGC 4472 lies at the center of a Virgo subcluster, 
its global gas temperature and density profiles are very
similar to those of other, more isolated ellipticals such as
NGC 4636.
We proceed in two stages.
First, the radial distribution of the total stellar and dark mass 
in NGC 4472 can be found by using the assumption of hydrostatic
equilibrium.
Then the usual evolutionary cooling flow equations can be 
solved over the Hubble time, using the known galactic potential,
to see how well the solutions can recover 
the gas and temperature variations observed in NGC 4472 today.

In Figure 4a we combine {\it Einstein} HRI (Trinchieri et al 1986)
and ROSAT HRI+PSPC (Irwin \& Sarazin 1996) data for the gas
(number) density and temperature distributions.
Then we fit these with the analytic curves shown and solved the 
hydrostatic equation for $M_{tot}(r)$ plotted with a solid line
in Figure 4b.
It is particularly gratifying 
that the hot gas is confined by the {\it stellar} potential
in the region $0.1~r_e \lta r \lta r_e$.
In this region the total mass determined from the x-ray gas 
agrees almost exactly with the 
de Vaucouleurs profile for the stellar mass $M_{*t}(r)$,
normalized by the stellar mass to light ratio determined by
van der Marel (1991).
Discounting a conspiracy of errors,
this excellent agreement implies a multitude of conclusions:
(1) the x-ray determined gas temperature is correct,
(2) van der Marel's stellar mass to light ratio is correct 
and {\it constant out to $r_e$}, and 
(3) ``mass dropout'' cannot be important in this large region of 
the cooling flow otherwise $M_{tot}$ would be less than
$M_*$ (Gunn \& Thomas 1996).
In $r \lta 0.1r_e$ we see that $M_{tot} < M_*$; this could 
be due to dropout but, if real, may be a result of large
self-generated magnetic fields that are expected to concentrate
near the galactic core (Mathews \& Brighenti 1997a).
Similar $M_{tot}(r)$ determinations 
for NGC 4649 and 4636 are discussed 
by Brighenti \& Mathews (1997a).

Using the total mass distribution $M_{tot}(r)$ from Figure 4b, 
we solved the cooling flow equations in the normal 
way, letting an initial gas-free galaxy evolve from $\sim 1$
Gyr to $t_n = 13$ Gyrs.
The objective is to reproduce the density and temperature 
profiles actually observed in NGC 4472.
The results of this calculation, shown as dotted lines
in Figure 5, are seen to differ significantly from the 
observations.
The computed gas density profile is too steep 
at {\it all} radii and 
the gas temperature is about half the observed value at 
$r \gta r_e = 8.6$ kpc.
In seeking agreement with the observations we varied the 
parameters and altered the assumptions in rather radical ways.
The usual mass dropout model only improved the slope of the 
gas density profile near the galactic center with no improvement
elsewhere and the deviation from the observed $T(r)$ was 
worsened.
Making the stellar mass loss rate or the thermal emissivity
of the hot gas increase (unphysically) with galactic radius 
had surprisingly little influence on the final cooling flows
at time $t_n$.
Increasing the supernova rate produced a huge iron abundance 
(four times solar) and only heated the gas modestly until
a galactic wind sets in.
Galactic winds are undesirable since $L_x$ drops far 
below the value observed for NGC 4472.

After these and other similar explorations, 
success in fitting the observed density and temperature was only
possible if the galaxy is filled with hot gas when the calculation 
is begun.
The dashed lines in Figure 5 
show the resulting interstellar gas
density and temperature 
when the calculation is begun with a total hot gas mass of 
$M_{hot} = 1.5 \times 10^{12}$ $M_{\odot}$ (much of which flows 
out during the subsequent evolution) all 
initially at temperature $T = 1.2 \times 10^7$ K.
At the end of the calculation 
when $t = t_n = 13$ Gyrs, 
only about 60 percent of the gas within $r_e$ comes from the stars.
If the gas from the stars is totally neglected throughout 
the evolution (dash-dotted 
line in Figure 5) the gas temperature is too large, but
reducing $\alpha_*(t_n)$ by a factor of two 
improves the fit somewhat (solid lines).

We conclude that the hot gas in galaxies like NGC 4472 cannot
have originated solely from mass loss from the currently observed 
stellar system.
Instead large masses of hot gas must have been present at early 
times and a significant fraction of this ancient 
gas is retained by present day ellipticals.
Evidently, all large ellipticals have this circumgalactic gas 
since $\langle T \rangle > T_*$ is generally observed 
(Davies \& White 1996).
Since the circumgalactic 
gas is hotter than the stars, its scale height 
in the galactic potential is also greater; this explains the 
relatively flat observed x-ray surface brightness distribution 
$\Sigma_x(R)$ and obviates the need for ``mass dropout.''
It is likely that the circumgalactic 
gas was heated by shocks in the 
secondary infall against the Hubble flow and by Type II supernova
explosions of massive stars.
The presence of old circumgalactic gas filling the distant
halo regions of present-day ellipticals will provide
important new information about the conditions that 
prevailed during the epoch of galaxy formation.

\acknowledgments

Our work on the evolution of hot gas in ellipticals is supported by
NASA grant NAG 5-3060 for which we are very grateful. In addition 
FB has been supported in part by Grant ASI-95-RS-152 from the Agenzia
Spaziale Italiana.


\clearpage

\figcaption[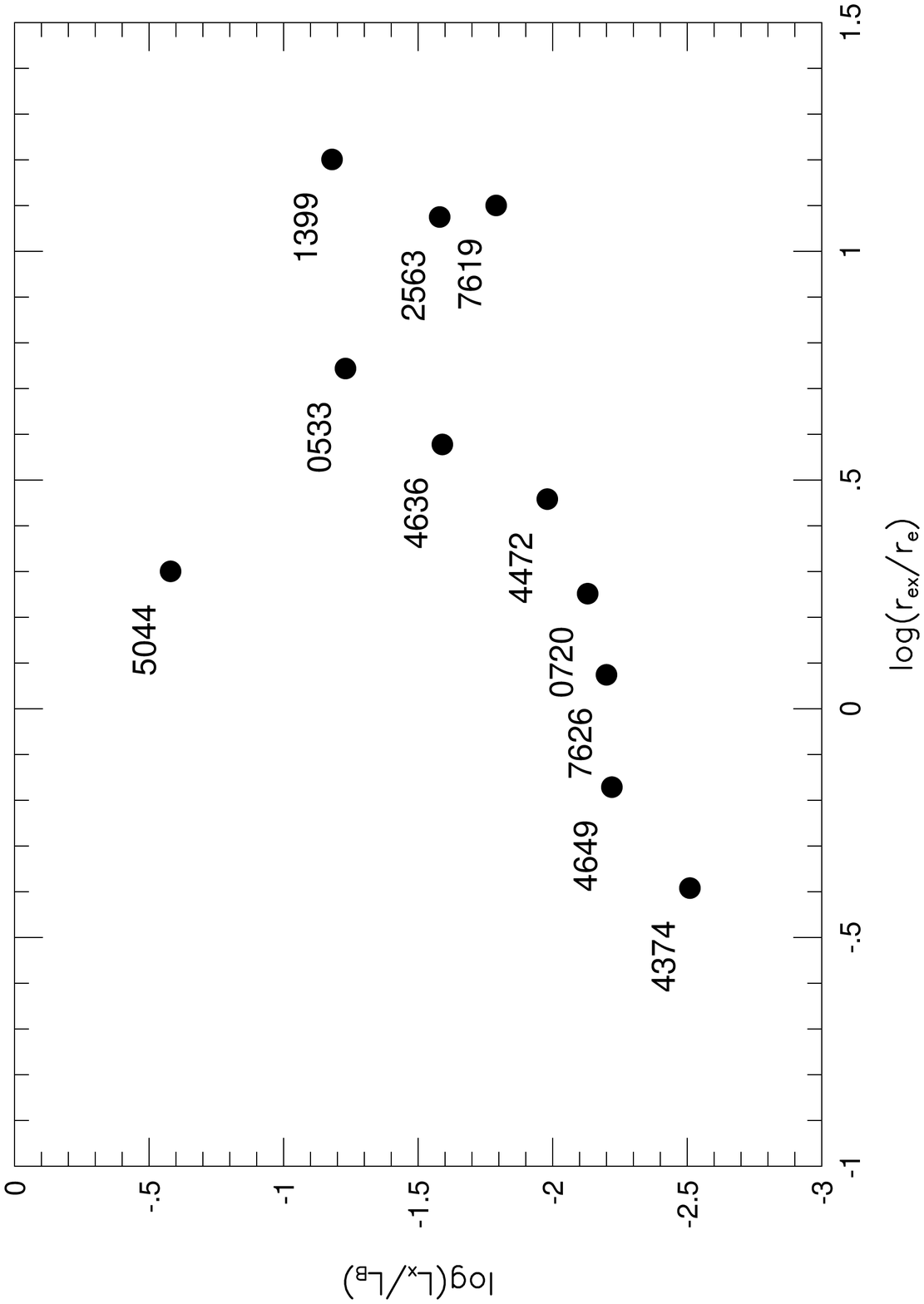]{
Variation of $L_x/L_B$ with $r_{ex}/r_e$ for eleven
well-resolved elliptical cooling flows.
\label{fig1}}

\figcaption[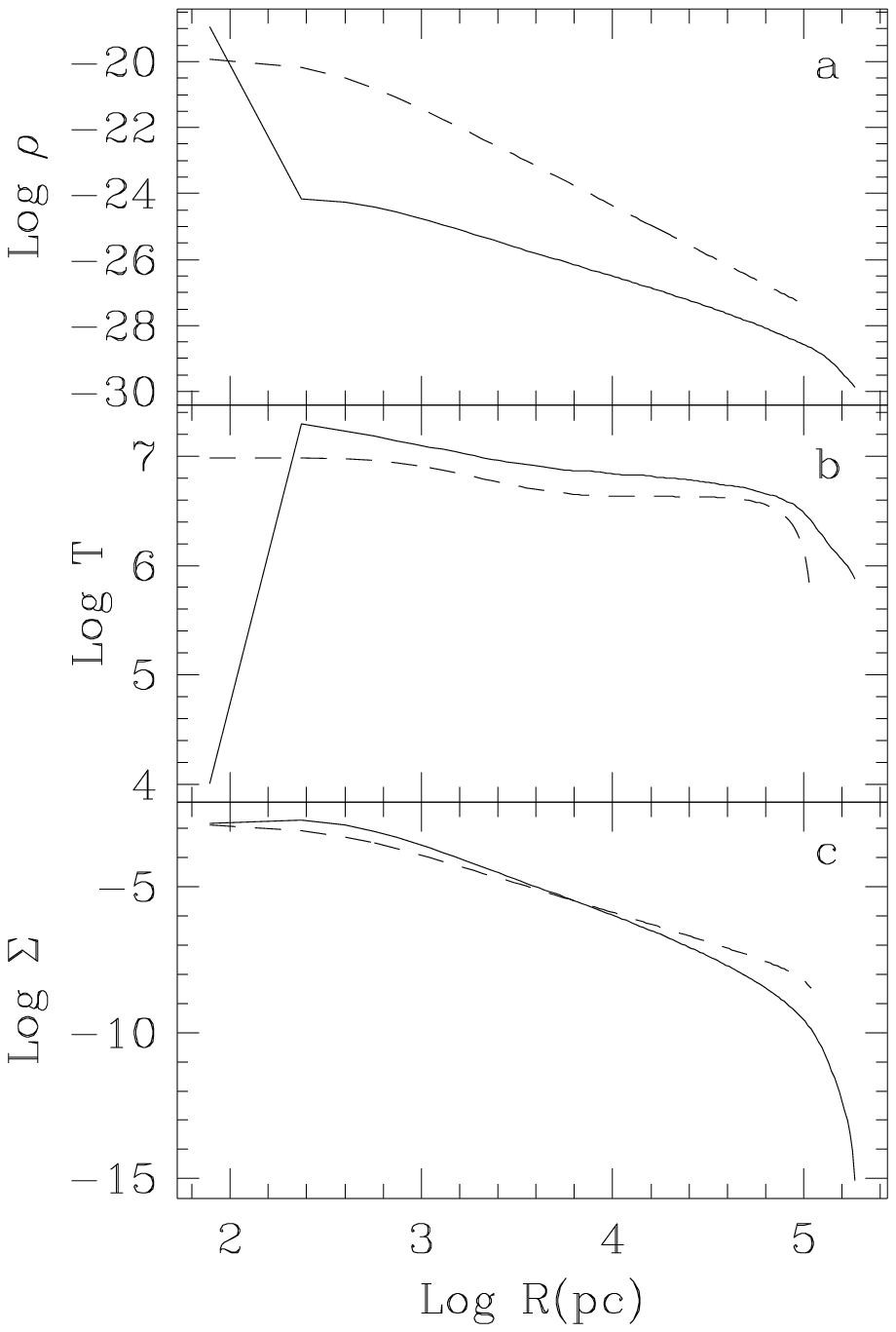]{
Cooling flow for an $L_B = 5 \times 10^{10} L_{B,\odot}$
elliptical at $t = 15$ Gyrs.
Density, temperature and surface brightness are shown for gas
({\it solid lines}) and stars ({\it dashed lines}) respectively.
The sudden transitions at small $R$ are artifacts of the
computational grid.
$\Sigma_x$ is computed for the band $0.5-4.5$ keV and
$\Sigma_*$ is arbitrarily normalized.
\label{fig2}}

\figcaption[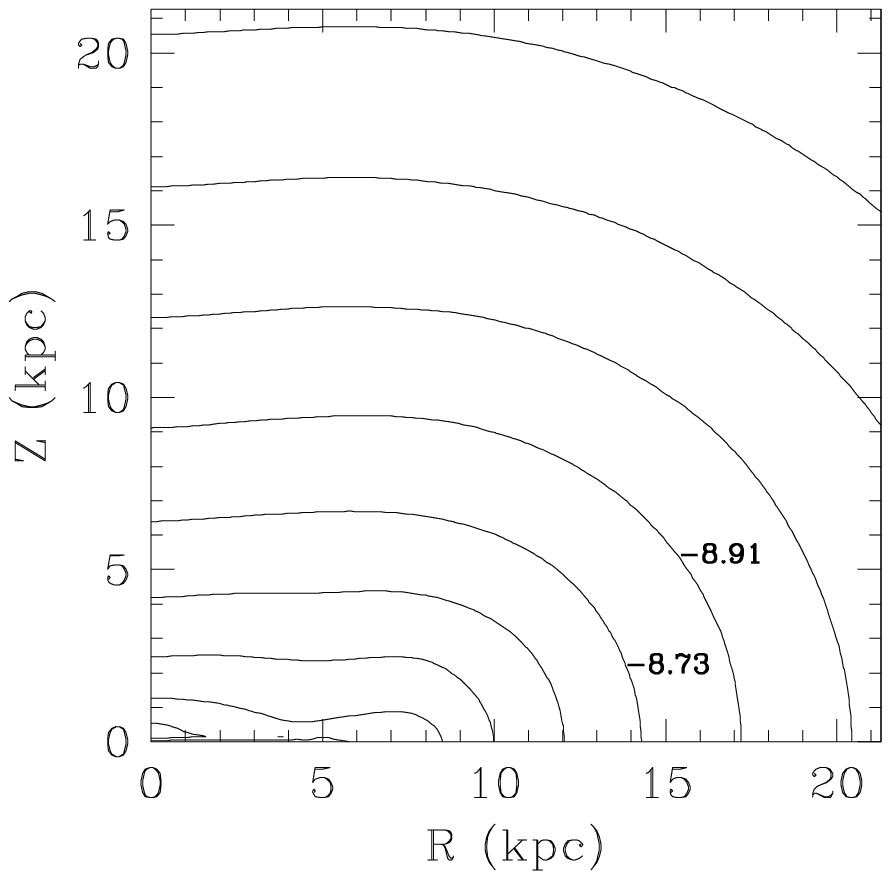]{
Equator-on x-ray surface brightness
distribution $\Sigma(R,z)$
in the central region of a slowly rotating
($v/\sigma = 0.39$) E2 galaxy
with $L_B = 5 \times 10^{10}$ $L_{B,\odot}$
after $t = 15$ Gyrs.
Two adjacent equally spaced contours are labeled with
$\log\Sigma(R,z)$ in ergs cm$^{-2}$ s$^{-1}$ for the
$0.4-4.5$ keV band.
\label{fig3}}

\figcaption[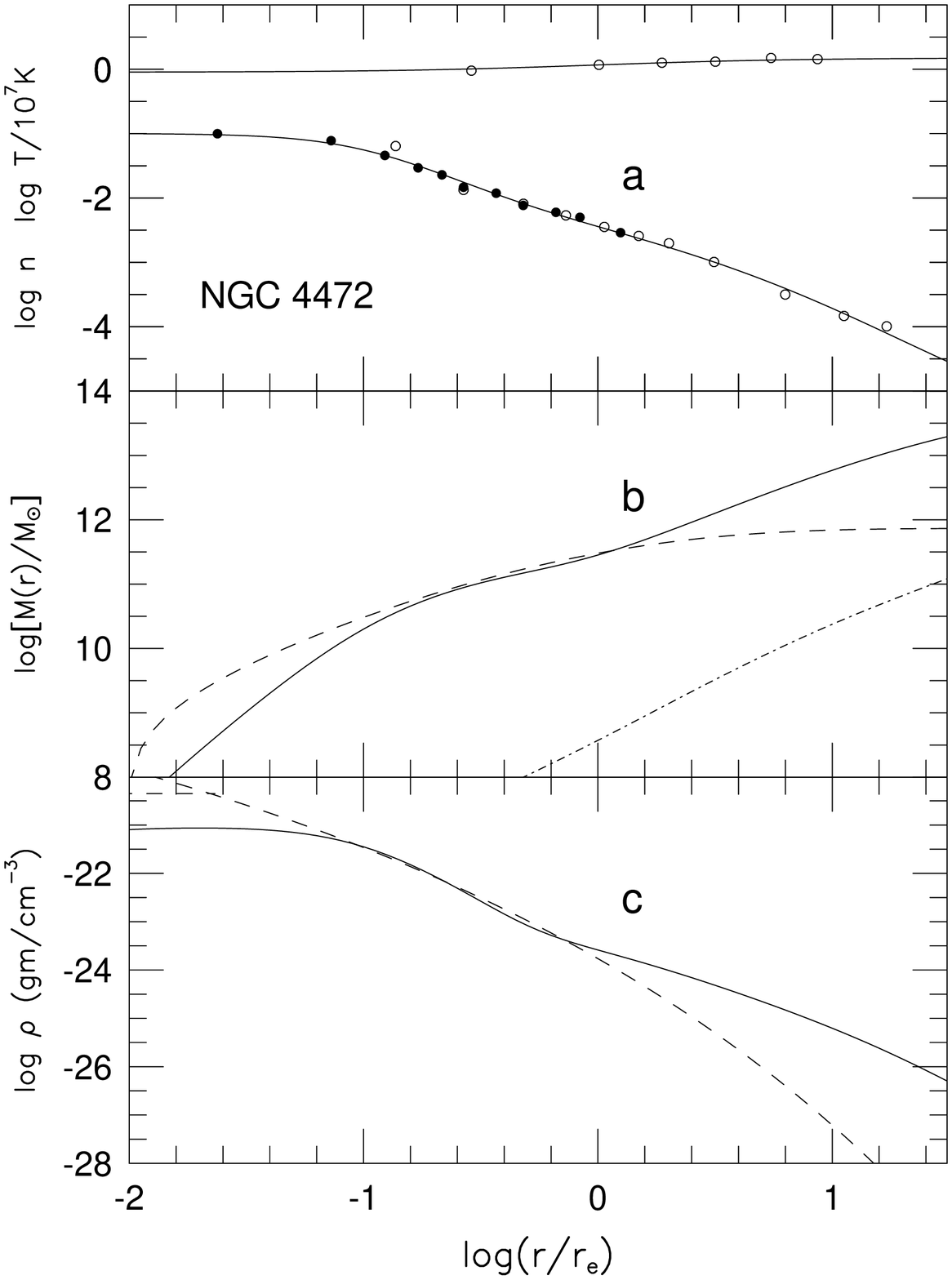]{
(a) {\it Einstein} HRI (filled circles)
and ROSAT (open circles)
observations of gas density and temperature in NGC 4472;
solid lines are analytic fits to data.
(b) Distribution of total mass $M_{tot}(r)$ (solid line),
stellar mass $M_*(r)$ (dashed line), and
hot gas mass $M_g(r)$ (dashed-dot line).
(c) Distribution of total mass density
$\rho_{tot}(r)$ (solid line)
and stellar density $\rho_*(r)$ (dashed line).
\label{fig4}}

\figcaption[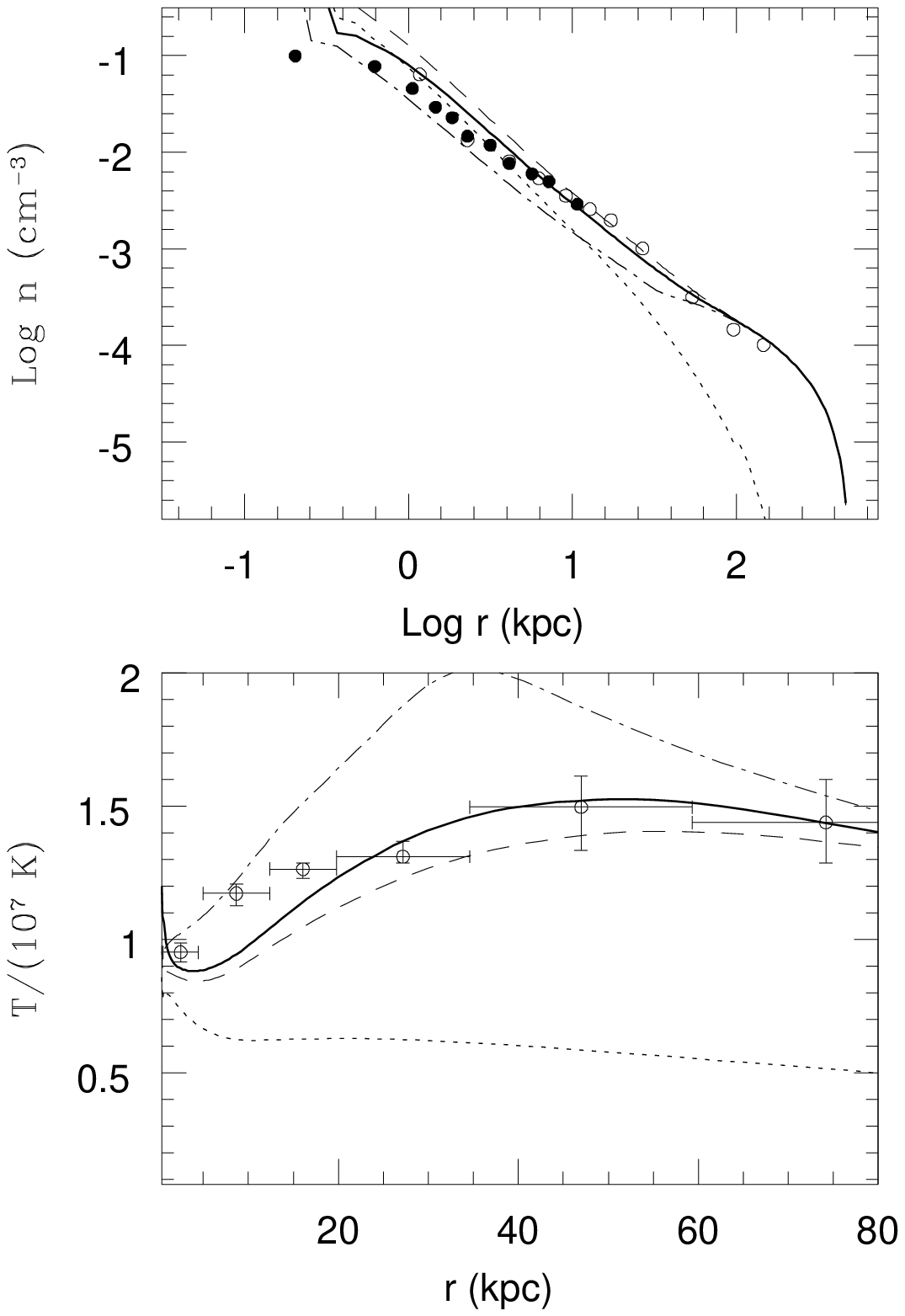]{
Gas density and temperature observations of NGC 4472
compared with density and temperature in
cooling flow models at $t = 13$ Gyrs with and without
additional gas at early times.
Dotted lines: cooling flow created only by stellar mass loss;
Dashed lines: same cooling flow with hot gas present at early times;
Solid lines: same cooling flow
with extra initial gas and $\alpha(t_n)$
reduced by 2;
Dot-dashed lines: same extra-gas cooling flow with $\alpha(t)=0$.
\label{fig5}}

\end{document}